\documentclass[prb,twocolumn,showpacs,amsmath,amssymb,preprintnumbers,superscriptaddress]{revtex4}
\usepackage{dcolumn}
\usepackage{bm}
\usepackage{graphicx}
\usepackage{subfigure}
\usepackage{color}
\usepackage{multirow,tabularx}

\begin{document}

\title{Space Charge Doping Induced Band Modulation in Mono- and Bi-layer Graphene: a nano-ARPES study}

\author{Imtiaz Noor Bhatti}\email{inbhatti07@gmail.com, imtiaz-noor.bhatti@synchrotron-soleil.fr}\affiliation{Synchrotron SOLEIL, L’Orme des Merisiers, Saint Aubin-BP 48 91192 Gif-Sur-Yvette Cedex, France}
\author{J. Avila}\affiliation{Synchrotron SOLEIL, L’Orme des Merisiers, Saint Aubin-BP 48 91192 Gif-Sur-Yvette Cedex, France}
\author{Z. Chen}\affiliation{Universit$\acute{e}$ Paris-Saclay, CNRS, Laboratoire de Physique des Solides, 91405, Orsay, France}
\author{A. Baron}\affiliation{Institut de Min$\acute{e}$ralogie, de Physique des Mat$\acute{e}$riaux et de Cosmochimie (IMPMC), Sorbonne Universit$\acute{e}$,75005 Paris, France}
\author{Y. Zhang}\affiliation{Institut de Min$\acute{e}$ralogie, de Physique des Mat$\acute{e}$riaux et de Cosmochimie (IMPMC), Sorbonne Universit$\acute{e}$,75005 Paris, France}
\author{A. Shukla}\affiliation{Institut de Min$\acute{e}$ralogie, de Physique des Mat$\acute{e}$riaux et de Cosmochimie (IMPMC), Sorbonne Universit$\acute{e}$,75005 Paris, France}
\author{P. Dudin}\affiliation{Synchrotron SOLEIL, L’Orme des Merisiers, Saint Aubin-BP 48 91192 Gif-Sur-Yvette Cedex, France}

\begin{abstract}
 Controlled modulation of electronic band structure in two-dimensional (2D) materials via doping is crucial for devices fabrication. For instance doped graphene has been envisaged for various applications like sensors, super-capacitors, transistors, p-n junctions, photo-detectors, etc. Many different techniques have been developed to achieve desired doping in 2D materials, like chemical doping, electrostatic doping, substrate doping, etc. Here, we have combined space charge doping with space and angle resolved photoemission (nano-ARPES), in order to directly observe the Fermi level modulation on micron-sized flakes of monolayer and bilayer graphene. The doping level can be tuned in a controlled manner, which allows us to directly observe the Fermi level tuning. In our experiment we successfully doped the graphene with p- and n-type carriers (holes/electrons) which are directly observed through band shift in ARPES measurements. The observed band shift is $\sim$250 meV for bilayer and $\sim$500 meV for monolayer graphene. The results from our experiment promote the space charge doping technique and nano-ARPES into other materials such as 2D semiconductors and superconductors, in order to directly observe the physical phenomena such as band gap transition and phase transition as function of carrier doping. 
\end{abstract}


\maketitle

Since the groundbreaking exfoliation of monolayer graphene in 2004 \cite{nov}, graphene and other two-dimensional (2D) materials, such as h-BN, MoS$_2$, SnS$_2$, WSe$_2$ and WTe$_2$ have attracted significant attention due to their remarkable properties.\cite{lev, rad, hua, wse1, wte1, wte2} As the first discovered 2D material, graphene has opened new frontiers in physics at the scale of single-atomic thickness, demonstrating exceptional electrical, mechanical, optical, and sensing capabilities.\cite{zhang, lee, nair, sch, tien} raphene’s unique combination of properties positions it as a potential replacement for conventional materials in existing applications. More importantly, its exceptional attributes collectively enable the development of groundbreaking technologies. For instance, the synergy of transparency, conductivity, and elasticity makes graphene ideal for flexible electronics, while the combination of transparency, impermeability, and conductivity is well-suited for transparent protective coatings and barrier films.\cite{nov2, ger1, ger2} The range of such innovative applications continues to expand, driven by the material's versatile property set.

Controlling carrier density is fundamental to tailoring the electronic properties of materials, enabling the exploration of various phase transitions. Doped graphene has been proposed for a wide range of applications, including electrochemical biosensing \cite{wha}, metal-free electrocatalysis \cite{gen}, lithium-ion batteries \cite{zho}, and supercapacitors \cite{jeo}. Precisely controlled n-type or p-type doping of graphene can further enable the fabrication of robust p-n lateral junctions \cite{bri} and vertical heterostructures \cite{will}, which hold significant potential for photodetection applications\cite{gab}. Further, direct visualization of chemical potential shifts and band structure changes controlled by the gate electric field has been demonstrated by ARPES study on a comples 2D device.\cite{ngu} Two widely employed approaches for tuning carrier density are gating and chemical doping, each offering distinct mechanisms to achieve this control. However, chemical doping has significant limitations: the doping density is difficult to control precisely, reversal requires high-temperature annealing, the random distribution of dopants introduces disorder, and the chemical interactions complicate accurate predictions of the electronic structure.\cite{tex} In contrast, electrostatic doping overcomes these challenges, offering precise control over carrier density without inducing disorder or chemical perturbations. Moreover, the carrier densities achieved via electrostatic doping are highly relevant for practical device applications.

\begin{figure*}
	\centering
		\includegraphics[width=14cm]{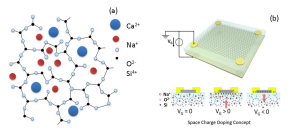}
	\caption{(color online) (a) Atomic arrangement in glass. (b) Schematic image of graphene on glass with doping conditions by applying gate voltage.}
	\label{fig:Fig1}
\end{figure*}

\begin{figure}
	\centering
		\includegraphics[width=8cm]{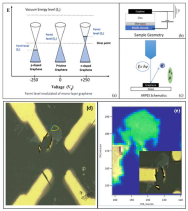}
	\caption{(color online) (a) Representation of effect of charge doping on Fermi level (b) Sample in device configuration, (c) ARPES measurement technique. (d) Optical image of sample with gold contacts. (e) Photoemission image of sample with inset same area in optical image.}
	\label{fig:Fig2}
\end{figure}

Manipulation of intrinsic electronic structures by electron or hole doping in a controlled manner in 2D materials is the key to control their electrical and optical properties \cite{cha}. ARPES is one of the most direct methods to study the occupied electronic states of solid. Electrostatic doping (back gating) is the most effective way to tune the carrier concentration in 2D materials, which has the advantage of ultraclean, inverseable and has ultrahigh doping range. Compare to the conventional electrostatic doping techniques that one needs to apply gate voltage in order to maintain the doping level, space charge doping technique has unique advantages: the efficient accumulation layer (electron doping) or depletion layer (hole doping) can be formed in glass substrate under gating voltage (VG) and temperature of $\sim$350 K, moreover the charge space can be quenched permanently after cooling the sample to room temperature or below. It means we can avoid of gating voltage once it is quenched \cite{as, ap}, for example, a semiconductor-metal-superconductor phase transition was discovered under the carrier concentration of N $\sim$10$^{15}$ cm$^{-2}$ in few-layer MoS$_2$ by space charge doping technique \cite{jb}.

Here in this work, we present ARPES study of space charge doped (electron/hole) mono-layer and bi-layer graphene on glass. We have demonstrate the shift in Fermi level around Dirac point with hole and electron doping. We have fabricated sample with mechanically exfoliated mono-layer and bi-layer graphene flakes which are transferred to the glass substrate. The graphene flakes are connected to gold electrode for grounding. ARPES measurements were performed at 70 K and in different doping conditions yield exciting results. We observed shift in Fermi level for both mono-layer and bi-layer graphene. Further, the space charge doping demonstrate non distinctive and reversible. moreover once space charge doping is created its maintain itself without need of applied voltage.  

%
Glass is an amorphous material that lacks a long range periodic crystalline structure. The 2D representation of a soda-lime silicate glass illustrates random atomic nature in Fig. 1. It is clearly seen that oxygen and silicon atoms form a SiO$_2$ network which is locally disrupted by Na$^+$ ions, where non-bridging oxygen ions serves as anions for Na$^+$. Heating the glass activates Na$^+$ ions mobility and it is possible to let them drift under the external applied electric field. Thus creates an accumulation or the depletion of Na$^+$ ions at the surface of the glass depending on the direction of field. A 2D material deposited on the surface of the glass is electrostatically n-doped by the accumulated Na$^+$ ions or p-doped by the uncompensated  O$^{2-}$ ions. Commercially available glass can be used for this purpose among them borosilicate glass (Na$^+$ ions concentration of $\sim$ 1.4 $times$ 10$^{18}$ cm$^{-3}$ ) and soda-lime glass (Na$^+$ ions concentration of $\sim$ 3.4 $times$ 10$^{18}$ cm$^{-3}$ ) are well tested. In this work we have used soda-lime glass.

\begin{figure}
	\centering
		\includegraphics[width=8cm]{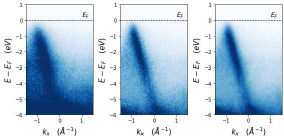}
	\caption{(color online) Variation of the electronic band structure around K point of monolayer graphene in undoped (centre) electron doped (right) and hole doped (left) is presented.}
	\label{fig:Fig3}
\end{figure}

\begin{figure}
	\centering
		\includegraphics[width=8cm]{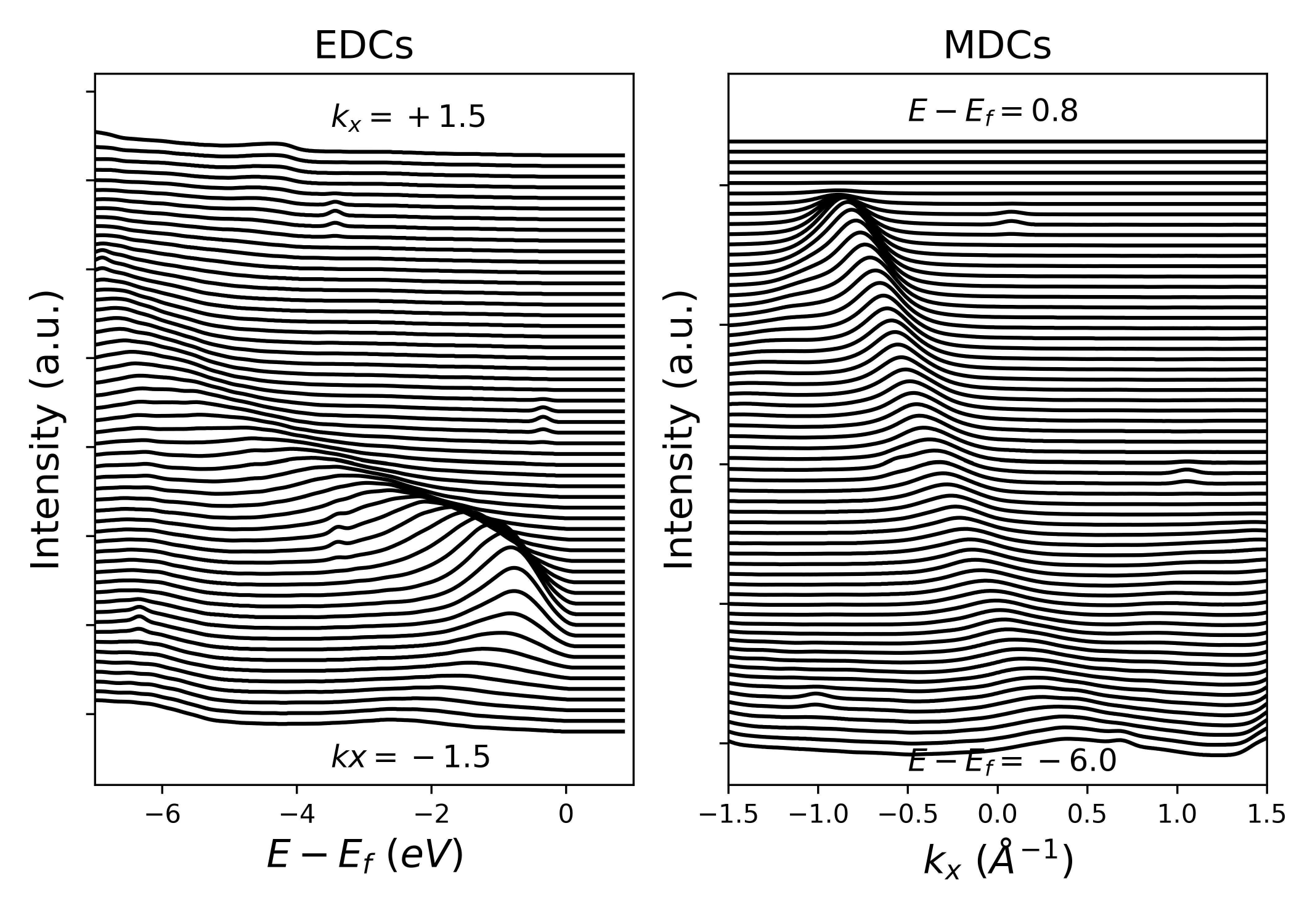}
	\caption{(color online) The energy distribution curves (EDCs) (left) and momentum distribution curves (right) for monolayer graphene}
	\label{fig:Fig4}
\end{figure}

Both samples of mono-layer and bi-layer graphene were produced using scotch micro-cleave method and next deposited onto the surface of soda-lime glass. The schematic of sample, measurement and expected doping effect on band is shown in Fig.2 b, c and a respectively. The graphene samples obtained with micro-cleave are typically only few tens of microns in size, as shown on the images of figure 2d blue encircle is monolayer flake where as yellow encircled area is bi-layer area of flake. To have the space-charged sample grounded it was electrically connected to the gold electrode using a piece of graphite (figure 2). Fig.2e. shows the optical image of sample with gold contacts and right shows the zoom photoemission spectroscopic image of same.
After introduction into vacuum of preparation chamber (base vacuum below 5$\times$10$^{-10}$ mbar) the samples were annealed at 160 $^o$C for 12 hours to preclean their surface after preparation in air. Immediately before the measurement of photoemission the samples were additionally annealed at 100 $^o$C for about 1 hour. After the cleaning by annealing, they were electron or hole doped by space charge doping technique, as described below. The band structure of monolayer graphene and bilayer graphene as function of doping was measured by angle-resolved photoemission with sub-micron spatial resolution (nano-ARPES, see [J.Avila et al., Journal of Electron Spectroscopy and Related Phenomena, 266, 2023, 147362] for details)\cite{avila}. The measurement was performed at temperature T = 70 K. The sample was well grounded, so local charging of glass substrate by scattered photons does not affect measurement.

\begin{figure}
	\centering
		\includegraphics[width=8cm]{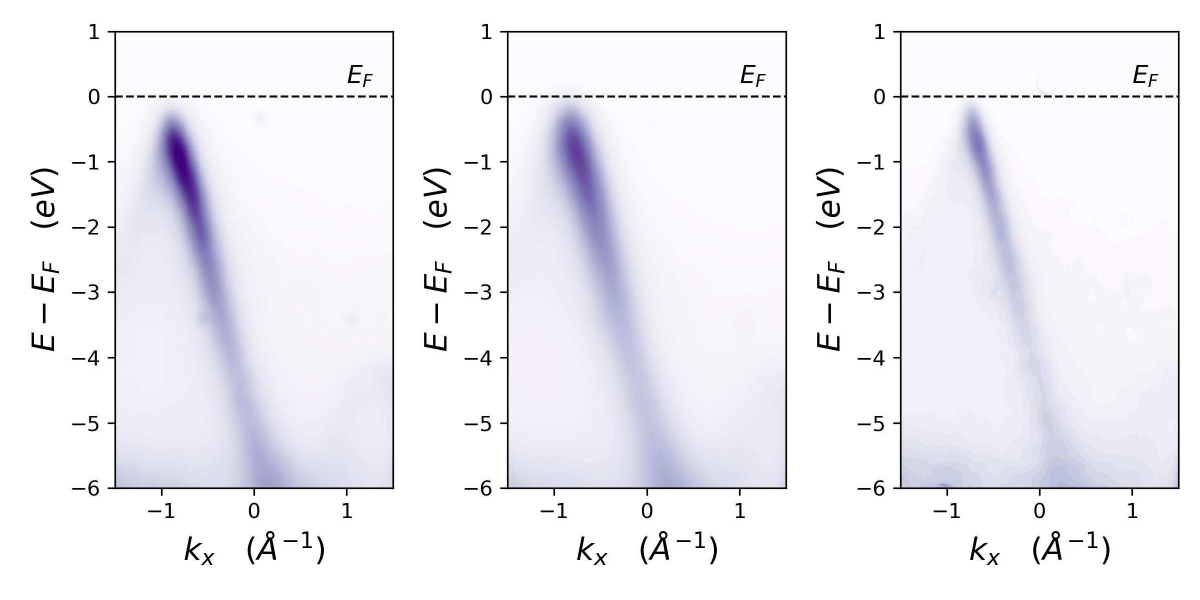}
	\caption{(color online) Variation of the electronic band structure around K point of bi-layer graphene in undoped (centre) electron doped (right) and hole doped (left) is presented.}
	\label{fig:Fig5}
\end{figure}

\begin{figure}
	\centering
		\includegraphics[width=8cm]{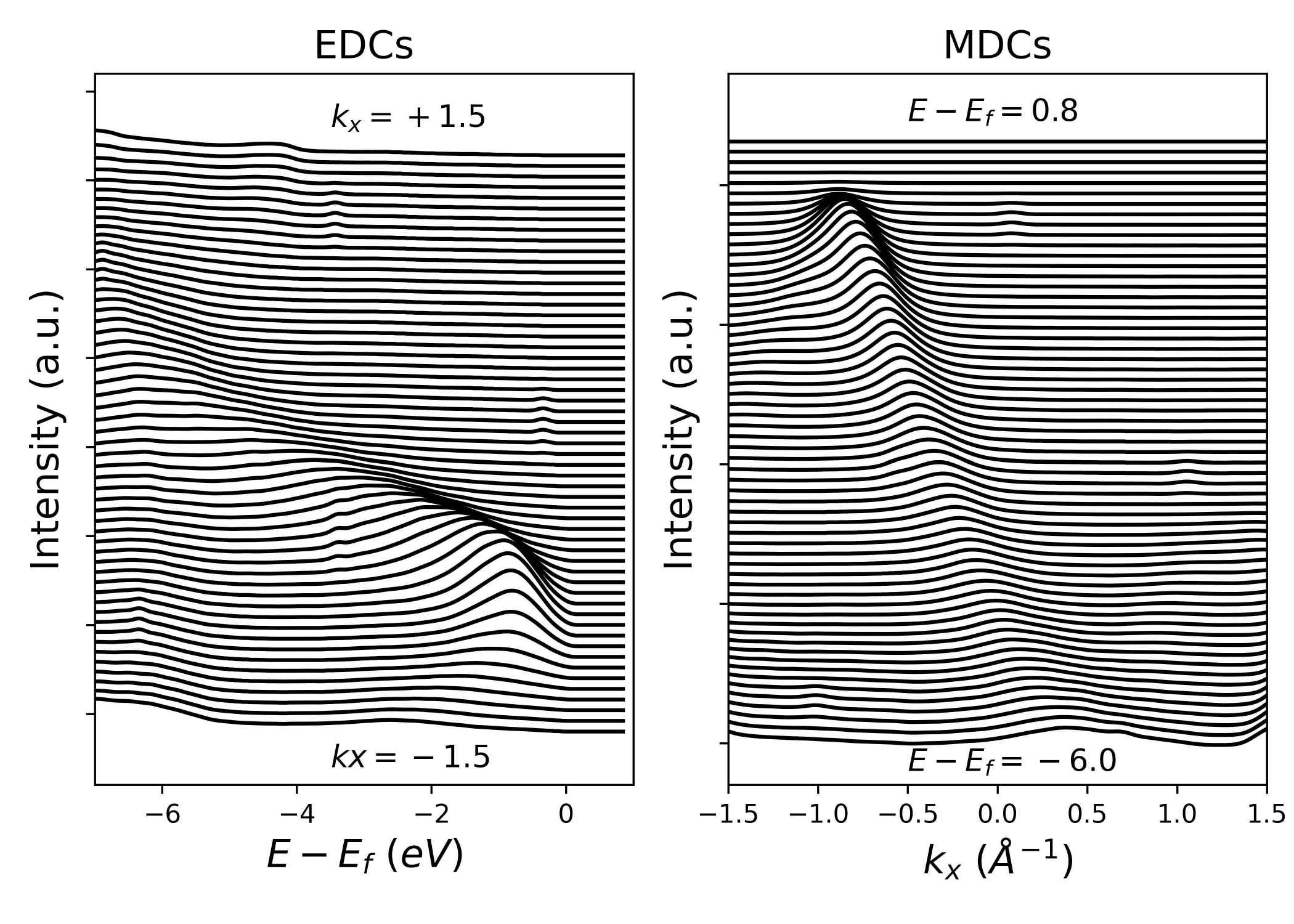}
	\caption{(color online) The energy distribution curves (EDCs) (left) and momentum distribution curves (MCDs) (right) for bi-layer graphene}
	\label{fig:Fig6}
\end{figure}

%
ARPES measurements were performed at SOLEIL Synchrotron. We have performed the measurements at 80 K with high intensity beam at 95 eV. The base pressure was kept below 10$^{-10}$ mbar throughout the measurement. The data was recorded using MBS analyser. ARPES measurement were carried on space charge doped mono-and bi-layer graphene. Charge doping of both polarities i.e. holes and electrons were done successfully. 

Fig.3 present the band around K point for mono layer graphene, the band structure for undoped sample is presented in the centre where as left and right are hole and electron doping respectively. The energy distribution curves (EDCs) and momentum distribution curves (MDCs) are shown in Fig.4. 

Similar measurements are performed for bilayer graphene, as shown in Fig.5. The band around K point is measured for undoped (center), electron doped, and hole doped case left and right of Fig. 5 respectively. The EDCs and MDCs for bilayer graphene is shown in Fig.6. 

We observe the shift of Fermi level with respective charge doping the Fig. 7 summarise the EDCs for comparison of undoped and doped  graphene left is for mono layer and right is for bi-layer. We observe a shift in fermi level with respect to undoped graphene in both the cases. The shift is around 500 meV for monolyer and 250 meV for bi-layer graphene. We successfully demonstrate the charge doping in graphene via space charge doping. This kind of charge doping is reversible and can be extended to future device applications.

\begin{figure}
	\centering
		\includegraphics[width=8cm]{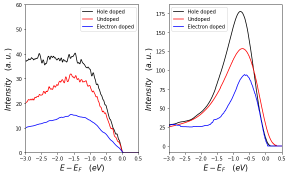}
	\caption{(color online) The EDCs for monolayer (left) and bi-layer (right) graphene presented for undoped and doped cases around K point.}
	\label{fig:Fig7}
\end{figure}

In conclusion, we sucessfully demonstrate ARPES study of space charge doping of graphene on glass. We successfully created space charges doping via high voltage and heating of glass followed by quenching below room temperature. Band structure shows broad bands for mono layer which may be attributed to amorphous nature of glass, however bilayer graphene does show sharp bands. We have measured fermi surface and band along -k direction for undoped and doped graphene. For mono- and bi-layer graphene shift of valance band maxima is seen at k, for mono layer the shift is $\sim$ 500 meV and for bilayer the shift is $\sim$ 250 meV. We have successfully doped both type of charges holes and electrons simply by reversing polarity of external voltage.

 Acknowledgement: Author Imtiaz Noor Bhatti acknowledge Synchrotron SOLEIL, France for financial support and Sorbonne University, Paris for facilities for the fabrication of samples.

Conflict of Interest: Authors have no conflict of interest to declare.

\end{document}